\documentclass[aps,prl,twocolumn,showpacs,letterpaper,superscriptaddress]{revtex4}
\usepackage{epsfig}
\usepackage{float}
\usepackage{color}
\usepackage{graphicx}
\usepackage{latexsym}
\usepackage{amsmath}
\usepackage{natbib}
\usepackage[english]{babel}

\newcommand{\dc}{\mathcal{D}}
\newcommand{\rv}{r }
\newcommand{\rvt}{\tilde{r} }

\newcommand{\rg}{\text{\tiny G}}
\newcommand{\rgt}{\tilde{\text{\tiny G}} }
\newcommand{\kv}{k }

\newcommand{\qvt}{\tilde{q} }

\newcommand{\vk}{v\kv}

\begin{document}
\title{\bf Efficient \textit{ab initio} calculations of bound and continuum excitons}

\author{Francesco Sottile}
\affiliation{European Theoretical Spectroscopy Facility (ETSF)}
\affiliation{Laboratoire des Solides Irradi\'es UMR 7642, CNRS-CEA/DSM, \'Ecole Polytechnique, F-91128 Palaiseau, France}

\author{Margherita Marsili}
\affiliation{European Theoretical Spectroscopy Facility (ETSF)}
\affiliation{Laboratoire des Solides Irradi\'es UMR 7642, CNRS-CEA/DSM, \'Ecole Polytechnique, F-91128 Palaiseau, France}
\affiliation{INFM-CNR-CNISM Dipartimento di Fisica, Universit\`a di Roma ``Tor Vergata'', Italy} 

\author{Valerio Olevano} 
\affiliation{European Theoretical Spectroscopy Facility (ETSF)}
\affiliation{LEPES - BP 166 - 25, avenue des Martyrs, 38042 Grenoble, France}

\author{Lucia Reining }
\affiliation{European Theoretical Spectroscopy Facility (ETSF)}
\affiliation{Laboratoire des Solides Irradi\'es UMR 7642, CNRS-CEA/DSM, \'Ecole Polytechnique, F-91128 Palaiseau, France}

\date{\today}

\begin{abstract}
We present calculations of the absorption spectrum of semiconductors and insulators comparing various approaches: 
(i) the two-particle Bethe-Salpeter equation of Many-Body Perturbation Theory; (ii) time-dependent density-functional 
theory using a recently developed kernel that was derived from the Bethe-Salpeter equation; (iii) a scheme that we 
propose in the present work and that allows one to derive different parameter-free  approximations to (ii). We show 
that all methods  reproduce the series of bound excitons in the gap of solid argon, as well as continuum excitons in 
semiconductors. This is even true for the simplest static approximation,
which allows us to reformulate the equations in a way such that the scaling of the calculations with number of atoms 
equals the one of the Random Phase Approximation.
\end{abstract}

\maketitle
Time-dependent density-functional theory (TDDFT) \cite{runge} is more and more considered to be a promising approach 
for the calculation of neutral electronic excitations, even in extended systems \cite{onida,tddftbook}. In linear response, 
spectra are described by the Kohn-Sham independent-particle polarizability $ \chi_0^{\tiny{KS}}$ and the frequency-dependent 
exchange-correlation (xc) kernel $f_{xc}$.
The widely used adiabatic local-density approximation \cite{zangwill,gross} (TDLDA), with its static and
short-ranged kernel,
often yields good results in clusters but fails for absorption spectra of solids. Instead, more sophisticated approaches
derived from 
Many-Body Perturbation Theory (MBPT) \cite{reining,sottile,adragna,marini,stubner}  have been able to reproduce, \textit{ab initio}, 
the effect of the electron-hole 
interaction in extended systems, not least thanks to an explicit long-range contribution \cite{deboeij,reining,botti}. 
The latter 
strongly influences spectra like optical absorption or energy loss, especially for relatively small momentum transfer.

Here we  show that this kernel is even able to reproduce the hydrogen-like excitonic series in the photoemission
gap of a rare gas solid. However the kernel has a strong spatial and frequency dependence, and its evaluation requires a 
significant amount of computer time. We therefore tackle 
 the question of a parameter-free, but quick TDDFT calculation of excitonic effects
in solids, which has been so far an unsolved problem, and show that a much more efficient formulation can indeed be achieved.
In particular we demonstrate how it is possible for a wide range of materials to obtain good absorption spectra including excitonic effects
with a {\it static} kernel leading in principle to a Random Phase Approximation (RPA)-like scaling of the calculation with the
number of atoms of the system.

Atomic units are used throughout the paper. The vectorial character of the quantities $r,k,G,q$ (where $k$ and $q$ are  vectors 
in the Brillouin zone, and $G$ is a reciprocal lattice vector) is implicit. Only transitions of positive frequency (i.e. resonant
contributions), which dominate absorption spectra, are considered throughout. 

\begin{figure}[t!]
\includegraphics[width=\columnwidth]{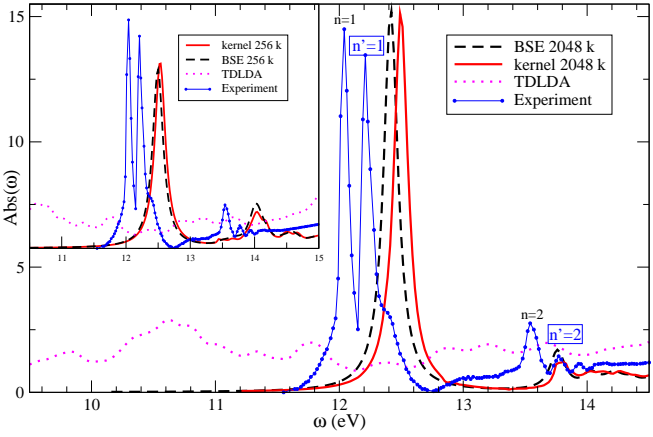}
\caption{Absorption spectrum of solid Ar. The BSE (dashed) and TDDFT using kernel of Ref.~\cite{sottile} (solid) are compared 
(only the $n'$ singlet exciton series) with experiment \cite{argonexp}. TDLDA is given by the points. Main panel: calculation 
with 2048 k-points. Inset: 256 k-points}
\label{argon}
\end{figure}

Let us first concentrate on the absorption spectrum of  solid argon. 
The low band dispersion, together with the small polarizability
 of the solid, conjures a picture where the
electron-hole interaction is  very strong and gives rise to a whole series of
bound excitons below the interband threshold. As in the optical spectra of other rare gas solids,
the first exciton is strongly bound (in argon by  $\sim 2$ eV), falling in
the class of  localized Frenkel \cite{frenkel} excitons. 
Closer to the continuum onset at 14.2 eV, one finds more weakly bound 
Mott-Wannier \cite{wannier} type excitons in a hydrogen-like series.  In the {\it ab initio} framework, such a complex spectrum is
typically described by the solution of the four-point (electron-hole) Bethe-Salpeter equation (BSE) 
\cite{hanke,strinati,onida}.
In Fig.\ref{argon} we show the optical spectrum 
of solid argon calculated within the BSE approach, and within TDDFT both using TDLDA 
\cite{details} and the MBPT-derived kernel \cite{sottile}.
The agreement of the BSE curve with experiment (line-circles) \cite{argonexp} (and with previous BSE calculations \cite{patterson}) 
is good, concerning both position and relative intensity of the first two peaks. It should be noted that the experiment shows double
peaks due to spin-orbit splitting, which is not taken into account in our calculations. The latter yields the singlet  excitons  that 
should essentially relate to the hole with $j=1/2$ and be compared with the $n'$ peaks. Besides the spin-orbit splitting, the
pseudopotential approximation as well as the construction of a static $W$ from LDA ingredients contribute to the remaining 
discrepancy with experiment. In spite of these limitations, the $n'=3$ peak can also be detected, although the 2048 k-points
used to calculate the spectrum are not sufficient to discuss it quantitatively, nor to describe the higher peaks. Instead,
the first two peaks require less k-points and, as can be seen in the inset, are already well reproduced with 256 k-points. 
In the following we therefore concentrate on these two structures and perform all calculations with 256 k-points.

The BSE impressively improves upon the TDLDA (dotted), 
which shows a structure-less 
broad curve, clearly missing the bound excitons. Instead, the kernel of Ref.~\cite{sottile} (full curve in Fig. \ref{argon})
leads to the same accuracy as the BSE, both for the Frenkel exciton and for the following structures. This demonstrates the
potential of the method and shows that the MBPT-derived kernel can be used to quantitatively predict the absorption spectra 
of a wide range of materials, including  the insulating  rare-gas solids. 

However,  the method is still computationally relatively heavy. Indeed in the MBPT-derived TDDFT approach, right as for the 
BSE two-particle Hamiltonian, one has to evaluate the matrix elements $F_{tt'}^{\text{\tiny{BSE}}}$ of the statically screened 
electron-hole Coulomb interaction $W$,
\begin{equation}
\label{eq:f-bse}
F_{tt'}^{\text{\tiny{BSE}}} \!=\! -2\pi\alpha\!\!\int\!\! dr_1 dr_2\,
\tilde{\Phi}^*_t(r_1,r_2)W(r_1,r_2)\tilde{\Phi}_{t'}(r_1,r_2)
\end{equation}
where the product $\tilde \Phi_t(r_1,r_2)=\phi_{vk}(r_1)\phi^*_{ck+q}(r_2)$ of two  KS wavefunctions $\phi$  is a generalized 
non local transition term; here 
$t$ is an index of transition with momentum transfer $q$, i.e. $t = \{vckq\}$, from valence $vk$ 
to conduction $ck+q$ states. $\alpha=2/(N_k\Omega_0)$ with $N_k$ number of k-points and
$\Omega_0$ volume of the unit cell.
The calculation of $F_{tt'}^{\text{\tiny{BSE}}}$ scales with the number of atoms $N_{at}$ as $N_{r} N_t^2 \sim N_{at}^5 $, 
where $N_r$ is the number of points in real space, and $N_t$ is the total number of transitions. 
Following Ref. \cite{sottile}, one then constructs an approximate kernel 
$f_{xc}^{\text{eff},{\mathcal A}} =  \chi_0^{-1} T^{\text{eff}}_{\mathcal A} \chi_0^{-1}$ with
\begin{equation*}
T^{\text{eff}}_{\mathcal A}(r,r',\omega)\!=\!\alpha\!\sum_{tt'}\!\frac{\Phi_t(r)}{(\omega+i\eta -\!\Delta E_t )}
F^{\text{\tiny{BSE}}}_{tt'}\frac{\Phi_{t'}^*(r')}{(\omega+i\eta -\!\Delta E_{t'} )}
\end{equation*}
where $\Phi_t(r_1)=\tilde \Phi_t(r_1,r_1)$ and
$\Delta E_t$ are differences between quasi-particle (QP) eigenvalues, since $f_{xc}^{\text{eff},{\mathcal A}}$ is
an approximation to the
 ``many-body''  kernel  $f_{xc}^{\text{mb}}$ that has to be used  in conjunction with an independent particle response 
function $\chi_0$ built with QP energies instead of Kohn-Sham (KS) ones as in pure TDDFT. This kernel simulates hence 
to good approximation the electron-hole interaction that is described by the BSE \cite{sottile}.

Even though this construction can be optimized \cite{marini} the method is at least an order of magnitude slower than an
RPA calculation. In the following we show how this problem can be overcome.

We concentrate on the irreducible polarizability $P$ that yields via the bare Coulomb interaction $v$ the reducible
polarizability $\chi$  from the matrix equation $\chi = P + P v \chi$, and the inverse dielectric matrix 
 $\epsilon^{-1} = 1+v\chi$. All quantities are functions of $q$ and frequency $\omega$, and matrices in $G,G'$.
Absorption spectra are then obtained from 
$\text{Abs}(\omega)= \displaystyle \lim_{q\to 0}\text{Im}\left\lbrace 1 / \epsilon_{00}^{-1}(q,\omega)\right\rbrace $. 
The polarizability $P$ is determined from the screening equation
$P  = \chi_0 + \chi_0 f_{xc}^{\text{mb}}P$.
In this equation we can now insert to the left and right of $f_{xc}^{\text{mb}} $  the identity  $1=XX^{-1}=X^{-1}X$, 
providing that $X$ is a non-singular function. This yields
\begin{eqnarray}\label{eq:tddftX}
P=\chi_0+\chi_0X^{-1}TX^{-1}P
\end{eqnarray}
where $T=Xf_{xc}^{\text{mb}}X$. We choose a matrix of the form $X=\alpha\sum_tg_t(\omega)\Phi_t(r)\Phi_t^*(r')$, 
where $g_t(\omega) $ is an arbitrary function. 
The term $T$ contains an explicit sum over matrix elements 
$F^{\text{\tiny{TDDFT}}}_{tt'} = 4\pi\alpha\int dr_1\,dr_2\,\Phi_t^*(r_1)f_{xc}^{\text{mb}}(r_1,r_2,\omega)\Phi_t(r_2)$  
in a basis of transitions $\Phi_t$, namely
\begin{equation}\label{eq:T}
T(r,r',\omega) \!=\! \alpha \sum_{tt'}g_t(\omega)\Phi_t(r)F^{\text{\tiny{TDDFT}}}_{tt'}g_{t'}(\omega)\Phi_{t'}^*(r').
\end{equation}
The exact $F^{\text{\tiny{TDDFT}}}_{tt'} $ is of course not known. However, in
the spirit of Refs. \cite{sottile,reining} we now replace the unknown  matrix elements $F^{\text{\tiny{TDDFT}}}_{tt'}$
with the BSE ones, given by Eq.(\ref{eq:f-bse}).  With this \textit{mapping}, $T$ is approximated as
\begin{widetext}
\begin{equation}
T \longrightarrow T^{\text{eff}}=\alpha\sum_{tt'}g_t(\omega)\Phi_t(r)\left[ \int dr_1\,dr_2\,\tilde \Phi_t^*(r_1,r_2)W(r_1,r_2)\tilde \Phi_{t'}(r_1,r_2)\right] \Phi_{t'}^*(r')g_{t'}(\omega) =\; X^3W\;^3\!X
\label{eq:Tmb}
\end{equation}
\end{widetext}
where we have defined a three-point right and left $X$ operator as: 
$X^3(r_1;r_2 r_{2'};\omega)=\alpha\sum_tg_t(\omega)\Phi_t(r_1)\tilde{\Phi}_t^*(r_2,r_{2'})$ and 
$^3\!X(r_1 r_{1'};r_2;\omega)=\alpha\sum_tg_t(\omega)\tilde{\Phi}_{t'}(r_1,r_{1'})\Phi_{t'}^*(r_2)$. 
Here it is important to 
underline that $F^{\text{\tiny{TDDFT}}}_{tt'}$ are constructed as matrix elements of the \textit{local} $\Phi_t(r)$, whereas 
$F_{tt'}^{\text{\tiny{BSE}}}$ are matrix element of the \textit{non-local}  $\tilde{\Phi}_t(r,r')$. In fact the  
{\it mapping} (\ref{eq:Tmb}) is not an exact operation, because  $F_{tt'}^{\text{\tiny{BSE}}}$ cannot  be expressed as 
a matrix element (between $\Phi_t$ and $\Phi_{t'}$) of a single $f_{xc}^{\text{mb}}$ for all $t,t'$ \cite{sottile,reining,sottile3}. 
Therefore $f_{xc}^{\text{eff}}=X^{-1}T^{\text{eff}}X^{-1} $ can be different from $f_{xc}^{\text{mb}}$, and  the quality 
of the resulting spectra will depend on the choice of $X$. 

If a certain freedom in the choice of $g_t$ can be exploited, one may find approaches that boost 
the computational efficiency with respect to $f_{xc}^{\text{eff},\mathcal A}$, i.e.  the expression of \cite{sottile}. 

In the following we will first illustrate, with the example of bulk Silicon and solid Argon, 
how different choices for $g_t(\omega)$ can lead to very similar spectra. 

We label with calligraphic letters the different choices $\mathcal{A,B,C,D}$ that stand for: 
\begin{equation}
\begin{split}
{\mathcal A}) \quad g_t(\omega)&=1/\left(\omega-\Delta E_t +i\eta\right),\quad \textrm{(i.e} \quad X=\chi_0\textrm{)}  \\
{\mathcal B}) \quad g_t(\omega)&=\text{Im}\left\lbrace 1/\left(\omega-\Delta E_t +i\eta\right) \right\rbrace\\
{\mathcal C}) \quad g_t(\omega)&=1/\Delta E_t
\qquad ; \qquad {\mathcal D}) \quad g_t(\omega)=1
\label{eq:choices}
\end{split}
\end{equation}

The first choice (${\mathcal A}$) defines nothing but the case $X=\chi_0$, as proposed in Ref.~\cite{sottile} and leading to 
$f_{xc}^{\text{eff},{\mathcal A}} $ above;
in the second case (${\mathcal B}$) only the imaginary part is taken from the denominator of the independent particle 
polarizability (very localized function in frequency); 
the cases (${\mathcal C}$) and (${\mathcal D}$) describe simple static choices for $g_t(\omega)$. 

\begin{figure}[t]
\includegraphics[width=\columnwidth]{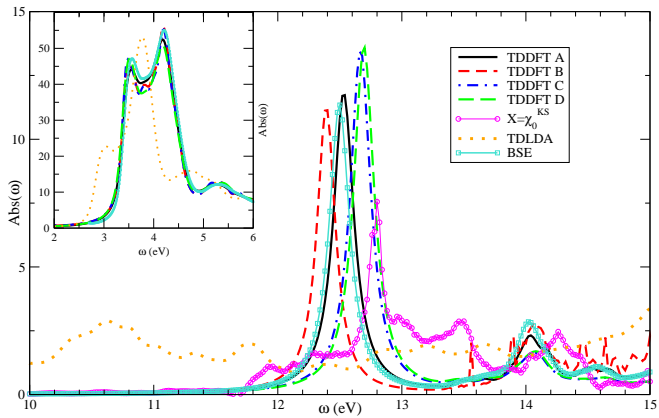}
\caption{Main: illustration of the different choices for Eq.(\ref{eq:choices}). In the inset: same choices for the absorption 
spectrum of Si. }
\label{kernels}
\end{figure}

The inset of Fig.\ref{kernels} shows the optical absorption of bulk Silicon calculated with the BSE and within TDDFT, 
using these mapping kernels ($\mathcal{A,B,C,D} $); the TDLDA result is also shown in order to emphasize the 
little differencies among the mapping kernels, 
compared to the huge improvements of ($\mathcal{A,B,C,D} $)  with respect to TDLDA.
The description of the optical absorption of Argon is a much more stringent test.
In Fig.\ref{kernels} 
we see that all the different kernels ($\mathcal A$ being slightly better than the others) are able to well reproduce the 
excitonic series and to strongly improve upon the TDLDA
result (dotted curve). This is especially surprising for choices ($\mathcal C$) and ($\mathcal D$): bound excitons have up
to now only been obtained using either the full, strongly frequency dependent kernel \cite{marini} or a frequency dependent 
long-range model $(\alpha + \beta \omega^2)/q^2$ \cite{botti2}, whereas a {\it static scalar} model can at the best yield
one single bound exciton, with largely overestimated intensity, by tuning appropriately two model parameters \cite{sottile3}. 
Our excellent results of Fig.\ref{argon} show, for the first time to the best of our knowledge, that  even a \textit{static}
parameter-free 
two-point kernel is able to reproduce 
a series of strongly bound excitons.

It is now crucial to understand and hence predict the performance of the
various choices, and to elucidate whether one can choose any possible $X\;$. 
To this aim  we start from the  four-point Bethe Salpeter equation 
$^4\!P=  ^4\!\chi_0+\;^4\!\chi_0W\;^4\!P $,
and contract the left and right indices. We obtain hence 
$P=\chi_0+ \chi_0^3 W\; ^3\!P,$
where we have defined a three-point right $\chi_0$ as \mbox{$\chi_0^3=\;^4\!\chi_0(r_1,r_1;r_2,r_{2'};\omega)$} 
and a three point left polarizability
$^3\!P=\;^4\!P(r_1,r_{1'};r_2,r_{2};\omega)$. Now, inserting the identity $1=\chi_0\chi_0^{-1}$, we obtain: 
\begin{equation}
P=\chi_0+\chi_0\chi_0^{-1} \chi_0^3W\;^3\!P.
\label{eq:BSE}
\end{equation}
On the other hand using the \textit{mapping} (\ref{eq:Tmb}) in eq.(\ref{eq:tddftX}), we obtain the approximate polarizability
\begin{equation}
P^{\text{eff}}=\chi_0+\chi_0X^{-1}X^3W\;^3\!XX^{-1}P^{\text{eff}}.
\label{eq:TDDFT}
\end{equation}

If TDDFT is to reproduce the BSE results, $P^{\text{eff}}$ resulting from (\ref{eq:TDDFT}) must be equal to $P$ of (\ref{eq:BSE}).

It should be noted that in principle the matrix $X$ can be chosen differently for the left and for the right side of $W$ in eq.
(\ref{eq:TDDFT}) and eq.(\ref{eq:tddftX}).
Concentrating first on the left side,
the choice $X=\chi_0$, i.e. $g_t(\omega)=1/(\omega + i\eta -\Delta E_t)$ recovers exactly the left side of $W$ in eq. (\ref{eq:BSE}).
\\
The right side is still to be optimised.  The comparison between (\ref{eq:BSE}) and (\ref{eq:TDDFT})
suggests to choose $X=P$. Of course this
is not the solution of the problem, since  i) $P$ is the quantity we are looking for and ii)  $P$ cannot be expressed as a sum over KS
transitions respecting the {\it ansatz} for $X$. 
Hence, one can only try to find a good guess. Again, $\mathcal A$, with $X=\chi_0$ seems a good choice. In fact, in a solid the 
joint density of states calculated in GW is very close to the density of transition energies calculated from the BSE; i.e. 
$\chi_0$ from GW and $P$ from the BSE have a very similar distribution of poles \cite{rohlfing}. 
Concerning the other choices, it is useful
to note that $P^{(3)}$ and $P$ have the same poles; the same statement holds for $X^{(3)}$ and $X$. If, in Eq.(\ref{eq:TDDFT}) 
the poles of $X^{(3)}$ cancelled with the zeroes of $X^{-1}$, and no new poles were introduced, one would just find the poles
of $P^{\text{eff}}$ in the right side of $W$ in Eq.(\ref{eq:TDDFT}), right as for $P$ in (\ref{eq:BSE}). 
However, $X^{-1}$ has poles that lie between the poles of $X$. These new poles are not problematic for energies in the continuum, 
but they can lead to
spurious structures when they appear isolated, i.e. in the bandgap. It turns out that this effect is particularly strong when the 
poles of $X$ are in the vicinity of the bound excitons. This is for example the case when one chooses $X=\chi_0^{KS}$
(i.e. $g_t(\omega)=1/(\omega+i\eta -\Delta E_t^{KS})$, $\Delta E_t^{KS}$ being the difference between KS
eigenvalues): indeed, the pink circles in Fig. \ref{kernels} show the  bad performance of that choice.. 

We have, in fact, verified that the spectra are generally very stable as long as we choose an $X$ that (i) either does not 
have {\it any} poles (static choices);
or (ii) has poles in the continuum (like $\chi_0$); or (iii) has poles at very low energies, much lower than all poles of $P^{\text{eff}}$.
This confirms that a \textit{wide range of choices for $X$} \textit{is indeed possible}. Moreover, this observation is valid for
a wide range of materials: we have performed the same test calculations for the prototype materials diamond and SiC, with similar conclusions.

As pointed out above, the aim is to avoid the unfavorable scaling of the calculations, determined essentially by the evaluation of  
$F^{\text{\tiny BSE}}_{tt}$ via Eq.(\ref{eq:f-bse}). 
Choice ($\dc$) is of course particularly simple and promising. 
In fact even when it is used as it is in (\ref{eq:Tmb}), the static choice $\mathcal{D}$ leads to a speedup with respect 
to choice $\mathcal A$ \cite{sottile}.
More importantly, it allows one 
to recombine the sums and integrals in Eq.(\ref{eq:Tmb}) in a more convenient way. The latter equation,
once  ($\dc$) is chosen, can in fact be written as
\begin{widetext}
\begin{equation}\label{eq:Tquick}
T^{\text{eff}}(q,\!\!\text{G},\!\!\text{G}') \!=\! -4\pi\alpha^2
\!\!\!\sum_{
\kv \tilde q \, \rgt\rgt' 
}
\!\!\!W_{\tilde \rg,\tilde \rg'}(\tilde q)
\!\!\!\int_{\Omega_0} \!\!\!\!\! d\rv d\rv' d\rvt d\rvt'   e^{-\imath\rg\cdot\rv} e^{\imath\rg'\cdot\rv'}
\!\!A_{\kv} (\rv,\rvt) B_{\kv-q} (\rvt'\!,\rv) A_{\kv+\qvt} (\rvt,\rv') B_{\kv+\qvt-q} (\rv'\!,\rvt')
e^{\imath\rgt\cdot\rvt} e^{-\imath\rgt'\cdot\rvt'}, 
\end{equation}
\end{widetext}
where $A_k(\rv,\rv')=\sum_{v}  u_{\vk}^*(\rv)  u_{\vk}(\rv')$,
$B_k(\rv,\rv')=\sum_{c}  u_{ck}^*(\rv)  u_{ck}(\rv')$, with the $u$'s representing the periodic
part of the KS wavefunction $\phi_{\vk}(\rv)=e^{-\imath \kv \cdot \rv}u_{\vk}(\rv)$.
$W_{\tilde \rg,\tilde \rg'}(\tilde q)$ is the reciprocal space Fourier transform of the statically screened
Coulomb interaction, with $\tilde q$ the difference between two k-points in the Brillouin zone. For $q\to 0$ we have the 
special case of vanishing momentum transfer $q$ (e.g. for optical absorption); (\ref{eq:Tquick}) is the  general formula valid for any 
 $q$ in order to treat also, e.g., electron energy loss or inelastic X-ray scattering. 

The scaling of Eq.\eqref{eq:Tquick} is in principle $N_{at}^4 $, but with a clearly dominant contribution given by the spatial integrals, which scales as
$N_{\rv}^3 \ln(N_{\rv}) < N_{at}^4 $ \cite{scaling}. Note that $N_{at}^4$ is the scaling of the construction
of $\chi_0$ itself \cite{efficient}. In other words, this formulation offers the possibility to determine absorption spectra 
including excitonic effects with a workload comparable to the RPA.

In conclusion, we have calculated the absorption spectra of solid argon, both by
solving the Bethe-Salpeter equation and by time-dependent density functional theory using a MBPT-derived \textit{mapping} kernel. 
Both methods yield results in good agreement with 
experiment and reproduce well positions and relative intensities of the peaks, with a drastic improvement over TDLDA results. 
We have then introduced a method that allows one to derive a variety of approximations for the TDDFT kernel; these can be used 
to tune computational efficiency while maintaining most of the precision of the original formulation. The method has been tested
for solid argon, silicon, diamond and silicon carbide. The good results, in turn, have allowed us to
propose a reformulation of the kernel (Eq.\eqref{eq:Tquick}) that leads  to  a TDDFT calculation with the same quality of the 
BSE, but with an RPA-like scaling $<N_{at}^4$, 
rather than $N_{at}^5$. This, we believe, can constitute a real breakthrough for practical applications where a low
computational effort - that characterizes TDDFT - and a precise description of many-body effects - like in the BSE - are required.

We are grateful for discussions with R. Del Sole and O. Pulci. This work was partially supported by the 
EU 6th Framework Programme through the NANOQUANTA Network of Excellence (NMP4-CT-2004-500198), ANR project XNT05-3\_43900, and by the 
Fondazione Italiana ``Angelo Della Riccia''. Computer time was granted by IDRIS (project 544).


\begin{thebibliography}{99}
\bibitem{scaling} The  approximation $W_{GG'}=W_{GG}\delta_{GG'}$, that is very often used in solids, moreover 
completely eliminates the $N_{at}^4$ terms leaving a pure 
 $N_{\rv}^3 \ln(N_{\rv}) $ scaling.
\bibitem{efficient} More efficient methods are available \cite{baroni} if one is interested 
only in one element of $\chi_{\rg,\rg'}$, and not in the whole matrix.
\end{thebibliography}
\end{document}